\documentclass[twocolumn]{aastex63}

\shorttitle{$DM_{\rm IGM}$ of FRB from IllustrisTNG and their cosmological applications}
\shortauthors{Zhang et al.}

\begin{document}

\title{Intergalactic medium dispersion measures of fast radio bursts estimated from IllustrisTNG simulation and their cosmological applications}

\correspondingauthor{F. Y. Wang}
\email{fayinwang@nju.edu.cn}

\author[0000-0001-8701-2116]{Z. J. Zhang}
\affiliation{School of Astronomy and Space Science, Nanjing University, Nanjing 210093, China}

\author[0000-0001-7542-5861]{K. Yan}
\affiliation{School of Astronomy and Space Science, Nanjing University, Nanjing 210093, China}

\author[0000-0002-9654-9123]{C. M. Li}
\affiliation{School of Astronomy and Space Science, Nanjing University, Nanjing 210093, China}

\author[0000-0001-6545-4802]{G. Q. Zhang}
\affiliation{School of Astronomy and Space Science, Nanjing University, Nanjing 210093, China}

\author[0000-0003-4157-7714]{F. Y. Wang}
\affiliation{School of Astronomy and Space Science, Nanjing University, Nanjing 210093, China}
\affiliation{Key Laboratory of Modern Astronomy and Astrophysics (Nanjing University), Ministry of Education, Nanjing 210093, China}

\begin{abstract}

Fast radio bursts (FRBs) are millisecond-duration radio transients and can be used as a cosmological probe. However, the dispersion measure (DM) contributed by intergalactic medium (IGM) is hard to be distinguished from other components. In this paper, we use the IllustrisTNG simulation to realistically estimate the $DM_{\rm IGM}$ up to $z\sim 9$. We find  $DM_{\rm IGM} = 892^{+721}_{-270}$ pc cm$^{-3}$ at $z=1$. The probability distribution of $DM_{\rm IGM}$ can be well fitted by a quasi-Gaussian function with a long tail. The tail is caused by the structures along the line of sight in IGM. Subtracting DM contributions from the Milky Way and host galaxy for localized FRBs, the $DM_{\rm IGM}$ value is close to the derived $DM_{\rm IGM}-z$ relation. We also show the capability to constrain the cosmic reionization history with the $DM_{\rm IGM}$ of high-redshift FRBs in the IllustrisTNG universe. The derived $DM_{\rm IGM}-z$ relation at high redshifts can be well fitted by a $tanh$ reionization model with the reionization redshift $z=5.95$, which is compatible with the reionization model used by the IllustrisTNG simulation. The $DM_{\rm IGM}$ of high-redshift FRBs also provides an independent way to measure the optical depth of cosmic microwave background (CMB). Our result can be used to derive the pseudo-redshifts of non-localized FRBs for $DM_{\rm IGM}<4000$ \ \ pc cm$^{-3}$.

\end{abstract}



\section{Introduction} \label{sec:intro}
Fast radio bursts (FRB) are millisecond luminous radio pulses with large dispersion measures (DMs), well in excess of the Milky Way contribution. There have been more than one hundred FRBs reported since \citet{2007Sci...318..777L} found the first one from archival data. Among them, only thirteen FRBs have been localized \citep{2017Natur.541...58C, 2019Sci...365..565B, 2019Natur.572..352R, 2019Sci...366..231P, 2020Natur.577..190M, 2020Natur.581..391M}. Apparently, FRBs can be divided into to two types, repeating and non-repeating ones and their host galaxy properties may be different.

The DM is defined as the column density of free electron density along a given line of sight (LoS). The observed DM is usually divided into several parts
\begin{equation}
\label{eq:DM}
DM = DM_{\rm{MW}} + DM_{\rm{halo}} + DM_{\rm{IGM}} + \frac{DM_{\rm{host}} + DM_{\rm{source}}}{1 + z}.
\end{equation}
In the above equation, $DM_{\rm MW}$ is the contribution of the interstellar medium in the Milky Way, which can be derived from the NE2001 Galactic free electron density model \citep{2002astro.ph..7156C} or the YMW model \citep{2017ApJ...835...29Y}. $DM_{\mathrm{halo}}$ is contributed by the free electrons in the Galactic halo. \citet{2019MNRAS.485..648P} found that the $DM_{\mathrm{halo}}$ is 50 pc cm$^{-3}$ \textless \ $DM_{\mathrm{halo}}$ \textless \ 80 pc cm$^{-3}$. Recently, \citet{2020ApJ...888..105Y} estimated that the mean $DM_{\rm halo}$ is 43 pc cm$^{-3}$, with a full range of 30-245 pc cm$^{-3}$. \citet{2019zhang} derived host contribution $DM_{\rm host}$ distributions of repeating and non-repeating FRBs with the IllustrisTNG simulation. They found that the distributions of $DM_{\mathrm{host}}$ can be well fitted by a log-normal function. For non-repeating FRBs, the median of $DM_{\mathrm{host}}$ is about $30 - 70$ pc cm$^{-3}$ in the redshift range $z=0.1-1.5$.  The $DM_{\mathrm{source}}$ depends on the central engine of FRBs. If FRB is generated by mergers of binary neutron stars \citep{2016ApJ...822L...7W, 2020ApJ...890L..24Z}, the value of $DM_{\mathrm{source}}$ is small \citep{2020ApJ...891...72W,Zhao2020}.

The DM contributed by intergalactic medium (IGM) is an important cosmological probe \citep{2014PhRvD..89j7303Z,2014ApJ...788..189G, 2016PhRvL.117i1301M, 2017A&A...606A...3Y, 2018A&A...614A..50W, 2018ApJ...856...65W, 2019MNRAS.484.1637J, 2019JCAP...09..039W, 2019ApJ...876..146L,2020ApJ...895...33W,Zhao2020a}. By assuming the cosmic reionization history, the value of $DM_{\mathrm{IGM}}$ can be derived theoretically from \citep{2003ApJ...598L..79I, 2004MNRAS.348..999I, 2014ApJ...783L..35D}
\begin{equation}\label{DM_{igm}}
DM_{\rm IGM}(z) = \frac{3c\Omega_{\rm b}H_0}{8\pi Gm_{\rm p}} \int^z_0\frac{ (1+z^\prime)f_{\rm IGM}(z^\prime)f_{\rm e}(z^\prime)}{E(z^\prime)}{\rm d}z^\prime,
\end{equation}
where $E(z)=H(z)/H_{0}$, $H(z)$ is Hubble parameter, $H_0$ is the Hubble constant, $m_p$ is the mass of proton, $\Omega_{\rm b}=0.0486$ is the density of baryons, and $f_{\rm IGM}$ is the fraction of baryon mass in the IGM. $f_{\rm e} = Y_{\rm H}X_{\rm e,H}(z)+\frac{1}{2}Y_{\rm He}X_{\rm e,He}(z)$. $Y_{\rm H}=3/4$ and $Y_{\rm He}=1/4$ are the mass fractions of hydrogen and helium, respectively. $X_{\rm e,H}$ and $X_{\rm e,He}$ are the ionization fractions of intergalactic hydrogen and helium.

Different from the previous theoretical investigations which use extragalactic DM with a homogeneous universe \citep{2003ApJ...598L..79I,2004MNRAS.348..999I}, \citet{2014ApJ...780L..33M} considered the effect of inhomogeneity with three models for halo gas profile of the ionized baryons. \citet{2015MNRAS.451.4277D}, \citet{2019ApJ...886..135P} and \citet{Zhu2020} studied $DM_{\rm IGM}$ with different cosmological simulations at low-redshift (\textit {z} \textless \ 2) universe.  \citet{2019MNRAS.484.1637J} used the Illustris simulation to estimate the $DM_{\rm IGM}$ and its scatters in the \textit {z} \textless \ 5 universe. However, the simulation accuracy can be improved with the latest IllustrisTNG simulation. Another advantage of IllustrisTNG is that it can provide the electron density directly instead of converting from the dark matter particle number density to baryonic matter density like \citet{2019ApJ...886..135P}. The complicated conversion may bring in extra uncertainties. Besides, the IllustrisTNG has a wide redshift range.

Considering these advantages, we choose the IllustrisTNG (the successor of Illustris) simulation which possesses both the high accuracy and large structures in the range 0 \textless \  \textit {z} \textless \ 20. Such a broad range of redshifts enables us to examine the prospect of constraining the cosmic reionization history with high-redshift FRBs. It also gives a chance to check whether FRBs can be a new type of standard candles besides supernovae, which is crucial for distance measurement.

In this paper, we use the IllustrisTNG simulation to study $DM_{\rm IGM}$ and their cosmological applications, especially at the high-redshift universe. The outline is as follows. An introduction to the IllustrisTNG simulation and our method to derive $DM_{\rm IGM}$ from the simulation are given in Section \ref{sec:method}. We show the result in Section \ref{sec:result}. The scenario of constraining the cosmic reionization history with FRBs is discussed in Section \ref{sec:dete}. We estimate the redshifts of non-localized FRBs with $DM_{\rm IGM}-z$ relation in Section \ref{sec:est}. Conclusions are given in Section \ref{sec:con}.

\section{Methods} \label{sec:method}

\subsection{Data access to IllustrisTNG}

\begin{figure*}[htb!]
	\centering
	\includegraphics[width=0.8\linewidth]{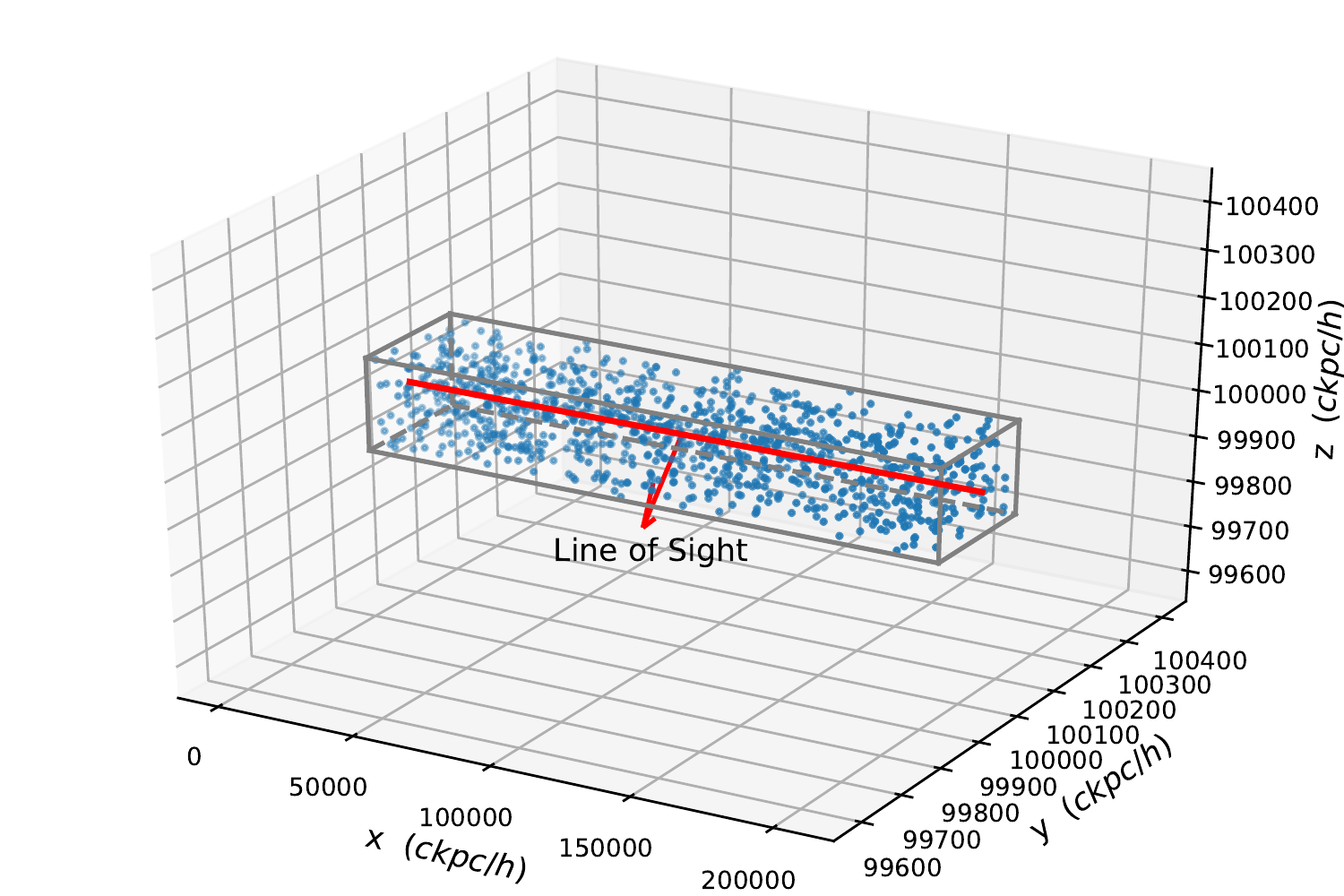}
	\caption{A schematic diagram of choosing LoS. There are 5125 pipes at each redshift as shown in this figure.}
	\label{pipe}
\end{figure*}

The IllustrisTNG project, a successor of Illustris, consists of three large volume, cosmological and gravo-magnetohydrodynamical simulations \citep{2018MNRAS.475..648P,2018MNRAS.475..676S, 2018MNRAS.475..624N, 2018MNRAS.480.5113M, 2018MNRAS.477.1206N}, such as TNG50, TNG100, and TNG300. The number represents the scale of simulations in the unit of cMpc (c for comoving similarly hereinafter). Each simulation contains several runs with different resolutions. The final results are stored in 600 HDF5 files. According to the cosmological principle, TNG300 is the best choice for our research and TNG300-1 is chosen among the three runs for its best resolution. There are 100 snapshots at different redshifts stored in each run and each snapshot has 15,625,000,000 Voronoi gas cells in total. Each cell corresponds to a particle and its physical parameters given by IllustrisTNG represent the whole cell. In the 100 snapshots, 20 snapshots are `full' and 80 snapshots are `mini' which only lack some particle fields. We use all the full snapshots at \textit{z} \textless \ 10 and several `mini' snapshots for better accuracy. The web-based JupyterLab Workspace and high-performance computing resources provided by IllustrisTNG  are used in this work.

\subsection{Dispersion of IGM}
For an FRB at \textit {z}$_{s}$, the $DM_{\rm IGM}$ can be written as
\begin{equation}\label{...}
DM_{\rm IGM}(z_s)=\int_{0}^{z_{s}} \frac{n_{\rm e}(z)}{1+z} {\rm d}l_{\rm prop},
\end{equation}
where $n_{\rm e}(z)$ is electron density in the comoving frame and ${\rm d}l_{\rm prop}$ is the differential proper distance. Then we use the redshift differential ${\rm d}z$ to express ${\rm d}l_{\rm prop}$ as
\begin{equation}\label{conv}
{\rm d}l_{\rm prop}=\frac {c}{H_{\rm 0}(1+z)E(z)}{\rm d}z,
\end{equation}
where
\begin{equation}\label{...}
E(z)=\sqrt{\Omega_{\rm m}(1+z)^3+\Omega_{\rm k}(1+z)^2+\Omega_{\rm \Lambda}}.
\end{equation}
So the $DM_{\rm IGM}$ can be rewritten as
\begin{equation}\label{dm_final}
DM_{\rm IGM}(z_{\rm s})=\frac {c}{H_0}\int_{0}^{z_{s}} \frac{n_{\rm e}(z)}{(1+z)^2E(z)}{\rm d}z.
\end{equation}
The cosmological parameters are taken as $\Omega_{\rm m}=0.3089$, $\Omega_{\Lambda}=0.6911$, and $H_0=67.74\ \rm{km\ s^{-1}\ Mpc^{-1}}$, which are the same as those used by the IllustrisTNG simulation \citep{2018MNRAS.475..648P}.

None of cosmological simulations provides continuous universe evolution, that's why snapshots exist. Therefore, equation (\ref{dm_final}) cannot be applied directly. In practice, \citet{2019MNRAS.484.1637J} and \citet{2019ApJ...886..135P} clipped and stacked the snapshots to construct the LoS. Here we solve the problem from another aspect. The basic idea is to convert the integral into summation.

If $n_{\rm e}(z_{\rm i})$ is available, the ${\rm d}DM_{\rm IGM}/{\rm d}z$ at a specific redshift ${z_{\rm i}}$ (${z_{\rm i}}$ = 0, 0.1, 0.2, 0.3, 0.4, 0.5, 0.7, 1, $\cdots$) can be derived by

\begin{equation}\label{dd}
\frac{{\rm d}DM_{\rm IGM}}{{\rm d}z}\bigg|_{z={z_{\rm i}}}= \frac {c}{H_0} \frac{n_{\rm e}(z_{\rm i})}{(1+z_{\rm i})^2E(z_{\rm i})},
\end{equation}
Then we use

\begin{small}
\begin{eqnarray}\label{}
&&DM_{\rm IGM}(z_{\rm i+1})= DM_{\rm IGM}(z_{\rm i})+\nonumber\\ &&\frac{1}{2}(\frac{{\rm d}DM_{\rm IGM}}{{\rm d}z}\bigg|_{{z_{\rm i}}}+\frac{{\rm d}DM_{\rm IGM}}{{\rm d}z}\bigg|_{{z_{\rm i+1}}})(z_{\rm i+1}-z_{\rm i})
\end{eqnarray}
\end{small}
to calculate the $DM_{\rm IGM}$ with the initial condition $DM_{\rm IGM}(z)\bigg|_{z=0}=0$.  Next subsection will introduce how to obtain the electron density.

\subsection{Calculations of $DM_{\rm IGM}$}

IllustrisTNG snapshot data is not organized according to spatial position. In order to obtain the average electron density along a given LoS, we use a traversal method on all the 15,625,000,000 particles and find the particles belonging to the given LoS. For computational simplicity, the LoS is chosen to parallel to the \textit{x} axis, which is similar to \citet{2019MNRAS.484.1637J}. Then we make 5125 square pipes with 200 ckpc/h ($\textit H_{\rm 0}\rm = 100\  \textit h\  \rm km\  s^{-1}\  Mpc^{-1}$) side in each snapshot and find the particles as well as necessary parameters (including \textit{Coordinates, Density, ElectronAbundance, GFM\_Metals and StarFormationRate}) in the pipes (see Figure \ref{pipe}). The 5125 pipes are chosen from different locations at 24 snapshots randomly and K-S test shows the sample size is representative. The electron density can be calculated by
\begin{equation}
n_{\rm e}(z)_{\rm prop}=\eta_{\rm e} X_{\rm H} \frac  {\rho }{m_{\rm p}} (1+z)^3,
\label{ne}
\end{equation}
where $\eta_{\rm e}$ is the electron abundance, $X_{\rm H}$ is the hydrogen mass abundance, $\rho$ is the density and $m_{\rm p}$ is mass of a hydrogen atom. The fourth parameter is the particle coordinate, which is used to select particles. The factor $(1+z)^3$ converts the density in the simulation comoving units into the proper units.

\begin{figure*}[htb!]
        \centering
        \includegraphics[width=1\linewidth]{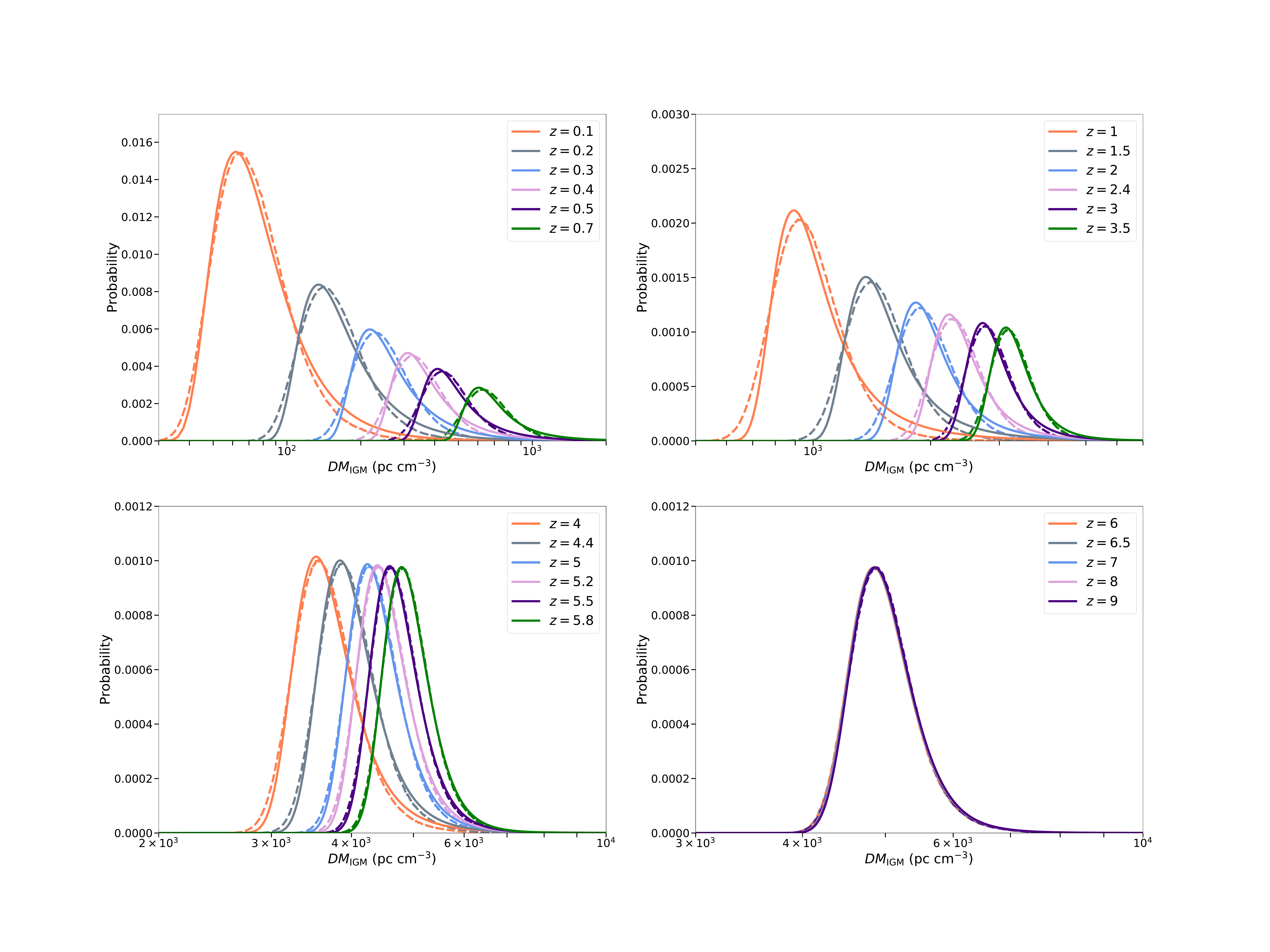}
	\caption{$DM_{\rm IGM}$ distributions from $z= 0.1$ to $9$. Dashed lines are $DM_{\rm IGM}$ distributions derived  from IllustrisTNG simulations and solid lines are the fitting results using equation (\ref{pIGM}). The best fitting parameters are shown in Table \ref{para}. At $z>6$, the distributions overlap as shown in panel (e), which indicates the universe do not reionize at such high redshifts.}
	\label{dis}
\end{figure*}

The equation (\ref{ne}) cannot be used in star-forming gas because the calculation is based on the `effective' temperature of the equation of state, which is not a physical temperature\footnote{\url{https://www.tng-project.org/data/docs/specifications/}}. Therefore, we use the parameter \textit{StarFormationRate} to exclude the star-formation gas. For comparison, star-forming particles are also excluded in \citet{2015MNRAS.451.4277D}. \citet{2019MNRAS.484.1637J} excluded gas cells belonging to haloes where dark matter density is 15 times larger than the cosmic critical density.

It is hard to get the electron density field along the LoS analytically, which requires to calculate the boundary of all the cells. While we know the cell to which any point on a LoS belongs. We divide the pipe into 10,000 bins along the \textit x axis and take the geometric center coordinates as the representation of bins. The distances between each bin and each particle in the pipe are calculated. We choose the nearest particle of each bin and assume the bin belongs to the cell of the chosen particle. We take an average of the electron density of 10,000 bins and put it into equation (\ref{dd}). As a result, 5125 ${\rm d}DM_{\rm IGM}/{\rm d}z$ are obtained at each redshift. Ten million $DM_{\rm IGM}-z$ relations are built by randomly selected ${\rm d}DM_{\rm IGM}/{\rm d}z$ from $z=0.1$ to 9.

\section{Result} \label{sec:result}
The redshifts of snapshots besides $z=0$ are shown in Table \ref{para}. The distributions of $DM_{\rm IGM}$ (from 0 to $z$ similarly hereafter) at different redshifts are shown in Figure \ref{dis}. The DM distributions are fitted with \citep{2014ApJ...780L..33M, 2019MNRAS.485..648P, 2020Natur.581..391M}:
\begin{equation}\label{pIGM}
p_{\rm IGM}(\Delta)=A\Delta^{-\beta}\exp[-\frac{(\Delta^{-\alpha}-C_0)^2}{2\alpha^2\sigma_{\rm DM}^2}], \Delta \textgreater 0,
\end{equation}
where $\Delta = DM_{\rm IGM}/<DM_{\rm IGM}>$, and $\beta$ is related to the inner density profile of gas in halos. We take $\alpha = \beta =3$, which is the same as \citet{2020Natur.581..391M}. $\sigma_{\rm DM}$ is an effective standard deviation. $C_0$, which can affect the horizontal position, is the remaining parameter to be fitted. The fitting results are shown in Table \ref{para}.

\begin{figure*}[htb!]
	    \centering
	\includegraphics[width=0.8\linewidth]{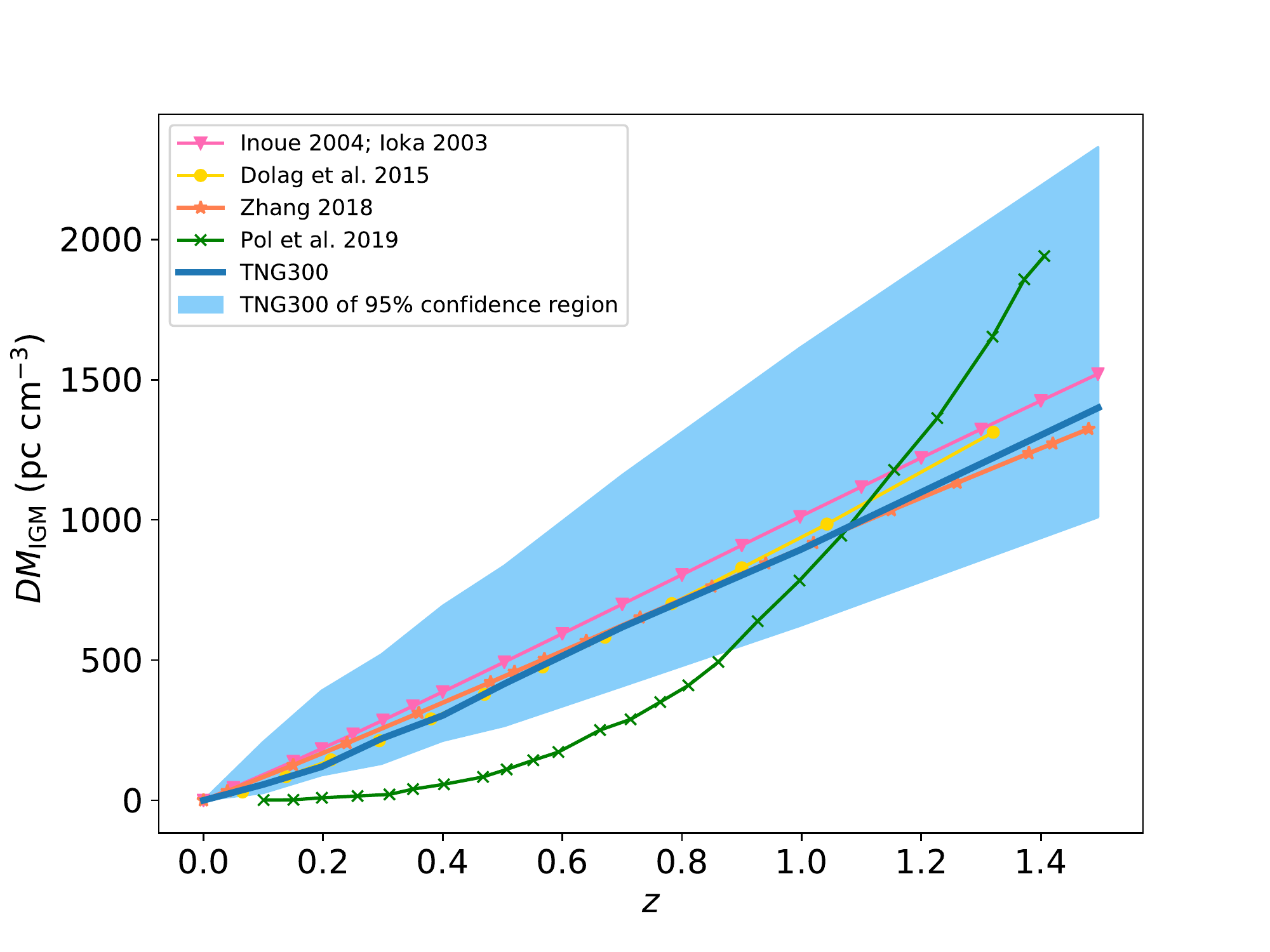}
	\caption{$DM_{\rm IGM}-z$ relation at low redshift ($z<1.5$). The blue line is our result from IllustrisTNG simulation and the blue shaded region is the 95\% confidence region. Other results are shown for comparison. The pink line with inverted triangle markers is taken from \citet{2003ApJ...598L..79I} and \citet{2004MNRAS.348..999I}. The dotted yellow line is taken from \citet{2015MNRAS.451.4277D}, the orange line with star marker is taken from \citet{2018ApJ...867L..21Z} and the green line with cross marks is \citet{2019ApJ...886..135P} uniform weighting result. Our result is consistent with other works except for the result of \citet{2019ApJ...886..135P}. However, if considering the large confidence level of \citet{2019ApJ...886..135P} (not shown in this figure), our result is consistent with theirs.}
	\label{c_other}
\end{figure*}

\begin{table}[htp!]
	\begin{tabular}{cccccc}
		\tableline
		$z$   & $A$        &   $C_0$      & $\sigma_{\rm DM}$   \\ \tableline
		0.1 & 0.04721 &  -13.17 & 2.554   \\
		0.2 & 0.005693 &  -1.008 & 1.118  \\
		0.3 & 0.003584  &  0.596 & 0.7043  \\
		0.4 & 0.002876 &  1.010  & 0.5158   \\
		0.5 & 0.002423 &  1.127  & 0.4306  \\
		0.7   & 0.001880  &  1.170   & 0.3595  \\
		1 &  0.001456   &  1.189  & 0.3044  \\
		1.5   &   0.001098 &  1.163   & 0.2609  \\
		2 & 0.0009672 &  1.162  & 0.2160  \\
		2.4   &  0.0009220&  1.142  & 0.1857  \\
		3 &  0.0008968& 1.119  & 0.1566  \\
		3.5   &  0.0008862&  1.104  & 0.1385 \\
		4 & 0.0008826 &  1.092  & 0.1233 \\
		4.4   &  0.0008827& 1.084  & 0.1134 \\
		5   & 0.0008834 & 1.076   & 0.1029 \\
		5.2   & 0.0008846 &  1.073  & 0.09918 \\
		5.5   & 0.0008863 & 1.070  & 0.09481 \\
		5.8  &  0.0008878 &  1.067  & 0.09072\\
		6 & 0.0008881 & 1.066&0.08971\\
		6.5  & 0.0008881 &1.066&0.08960\\
		7 &  0.0008881 &1.066&0.08952\\
		8 & 0.0008881 &1.066&0.08944\\
		9 & 0.0008881 &1.066&0.08941\\
		\tableline
	\end{tabular}
	\caption{Redshifts of the snapshots and fitting parameters of the $DM_{\rm IGM}$ distributions in Figure \ref{dis}. \label{para}}
\end{table}
It's obvious that the asymmetrical distributions have long tails at high $DM_{\rm IGM}$, so we choose the most probable value for analysis, which is also used by \citet{2015MNRAS.451.4277D} and \citet{2019ApJ...886..135P}. We find $DM_{\rm IGM}(z=1) = 892^{+721}_{-270}$ pc cm$^{-3}$ (errors represent 95\% confidence level). \citet{2003ApJ...598L..79I} and \citet{2004MNRAS.348..999I} predicted $DM_{\rm IGM}(z=1) \sim 1200$ pc cm$^{-3}$. \citet{2018ApJ...867L..21Z} predicted  $DM_{\rm IGM}(z=1) \sim 855 \pm 345 $ pc cm$^{-3}$.  \citet{2019MNRAS.484.1637J} found $DM_{\rm IGM}(z=1) \sim 905 \pm 115$ pc cm$^{-3}$ (errors represent $1\sigma$ standard deviation). \citet{2019ApJ...886..135P} derived $DM_{\rm IGM}(z=1) \sim 800^{+7000}_{-170}$ pc cm$^{-3}$ with uniform weighting and $DM_{\rm IGM}(z=1) \sim 960^{+350}_{-160}$ pc cm$^{-3}$ with weighting by matter distribution (errors represent 95\% confidence level). In Figure \ref{c_other}, we show these above $DM_{\rm IGM}-z$ relations. Our result is shown as blue solid line with 95\% confidence region (blue region). The $DM_{\rm IGM}$ estimated in our work is consistent with others within 95\% confidence level including the one derived by \citet{2019ApJ...886..135P}. The difference between \citet{2019ApJ...886..135P} and our result may be caused by the conversion from the dark matter number density to the free electron density in the MareNostrum Instituto de Ciencias del Espacio Onion Universe simulation. A non-negligible systematic error may be from the different cosmological parameters used by these simulations. For example, Illustris uses the cosmological parameters from WMAP-9 measurements \citep{2014Natur.509..177V}, while IllustrisTNG uses those from \textit{Planck} \citep{2018MNRAS.475..648P}.

\begin{figure*}[htb!]
		    \centering
	\includegraphics[width=0.8\linewidth]{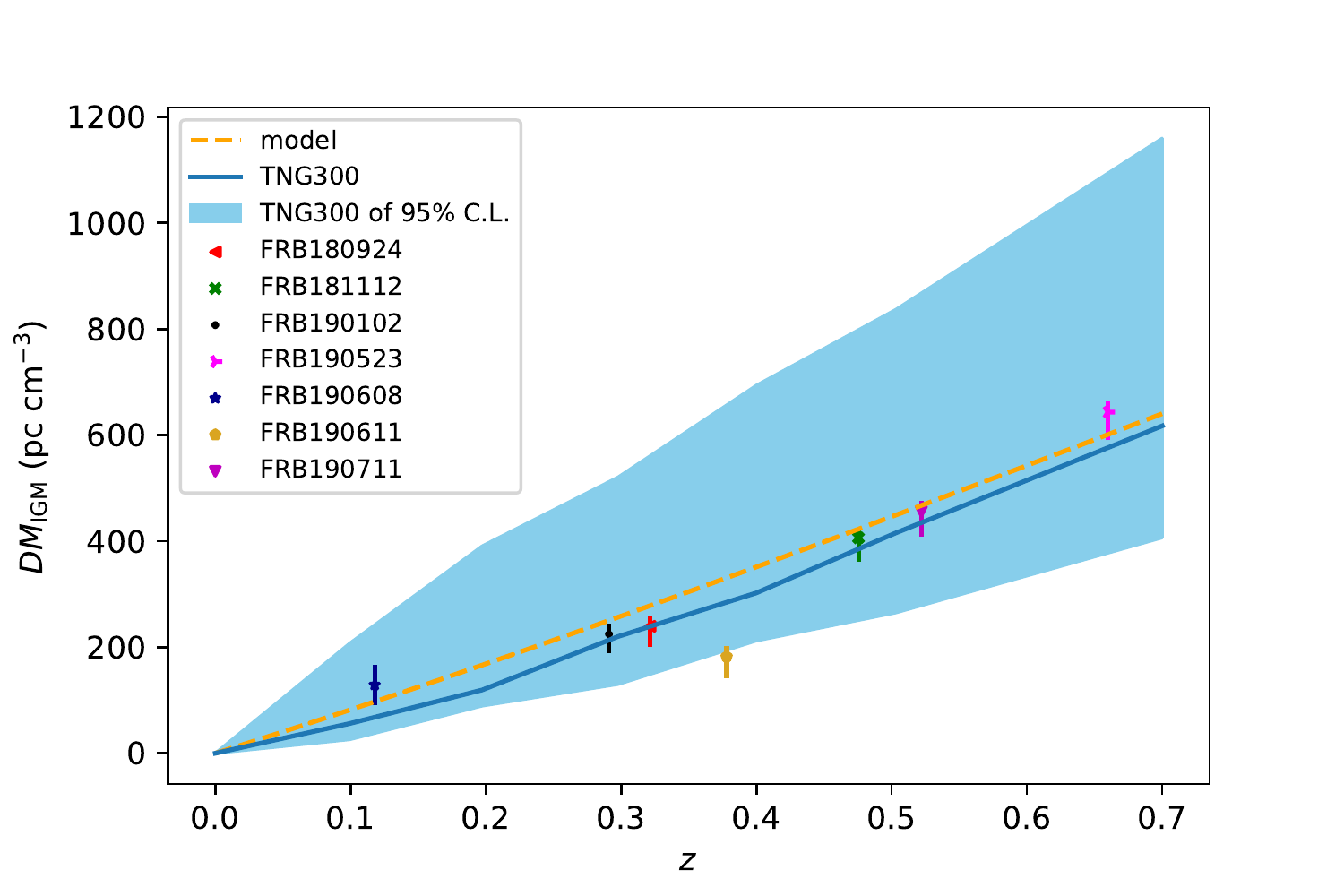}
		
	\caption{$DM_{\rm IGM}-z$ (Macquart) relation from IllustrisTNG simulation and several localized FRBs. The orange dashed line is the model of equation (\ref{DM_{igm}}). We take $f_{IGM}=0.82+0.08z/0.9$, $f_{\rm e} = 7/8$, and $X_{\rm e,H}=X_{\rm e,He}=1$. The blue line is our result derived from TNG300 and the blue shaded region is the 95\% confidence level. The two lines differ because the ionized fraction and IGM baryon fraction are different between the model and the simulation. We also show the $DM_{\rm IGM}$ value of several localized FRBs. In calculation, the $DM_{\rm MW}$ value is derived with the NE2001 model and $DM_{\rm host}$ is taken from  \citet{2019zhang} except FRB 190608 (taken from \citet{2020arXiv200513158C}).}

	\label{c_obs}
\end{figure*}

We also compare our result with several localized FRBs \citep{2019Sci...365..565B, 2019Natur.572..352R, 2019Sci...366..231P, 2020Natur.581..391M} in Figure \ref{c_obs}. The blue line is our result derived from TNG300 and the blue shade region is the 95\% confidence region. The theoretical one from equation (\ref{DM_{igm}}) is shown as dash line for the cosmological parameters from Planck \citep{2016A&A...594A..13P}. The $DM_{\rm host}$ value is adopted from \citet{2019zhang}. Based on the host galaxy observations, \citet{2019zhang} calculated the $DM_{\rm host}$ of repeating and non-repeating FRBs from the IllustrisTNG simulation. They found $DM_{\rm host}=32.97(1+z)^{0.84}$ pc cm$^{-3}$ for non-repeating FRBs. For FRB 190608 we adopt its $DM_{\rm host}=137\pm43$ pc cm$^{-3}$ from observations \citep{2020arXiv200513158C}. The $DM_{\rm MW}$ value is derived with the NE2001 model \citep{2002astro.ph..7156C}. We take $DM_{\rm halo}=50$ pc cm$^{-3}$ for all FRBs. The contributions from FRB sources are ignored. All the derived $DM_{\rm IGM}$ are compatible with our result.

\section{The scenario of constraining the cosmic reionization history} \label{sec:dete}
High-redshift FRBs can be used to probe the cosmic reionization history and we test the expectation with TNG300 simulation. The universe is mostly neutral and non-transparent after the recombination \citep{1968ApJ...153....1P,1969JETP...28..146Z,2000ApJS..128..407S}. The transition from the neutral to ionized universe called reionization, which is the first chance to detect the universe evolution after the end of cosmic dark ages. Reionization history is a frontier and challenging area in cosmology \citep{2013fgu..book.....L}, because it is hard to detect now. Usually, it is believed that the reionization of hydrogen occurs between $z \sim 12$ and $z \sim 6$. For helium the range is between $z \sim 6$ and $z \sim 2$ \citep{2016A&A...596A.108P}. Some works have proposed to use the FRBs to constrain the reionization history \citep{Zheng2014,2016JCAP...05..004F,2020PhRvD.101j3019L,DaiXia2020,Bhattacharya2020}.

It is still difficult to localize non-repeating FRBs even at low redshifts. The capability of localization is improving. \citet{2020arXiv200811738L} demonstrated that the blind interferometric detection and localization of non-repeating FRBs by the CHIME and the CHIME Pathfinder can be down to about milliarcsecond precision. In future, they will observe thousands of single FRB events to this precision. However, considering the offsets which have been observed in FRBs, spectroscopy should be applied for high-redshift FRBs besides radio interferometer. To confirm the redshift, optical/infrared observations should also be applied. Finding high-redshift quasars is a challenging task at present, let alone the high-redshift FRB hosts which may be faint.

Combining the high burst rate and the construction of next generation telescopes, it is hopeful to constrain the reionization history with FRBs. The estimated $\gtrsim 10^3$ day$^{-1}$ rate of FRBs \citep{2019A&ARv..27....4P} would provide abundant chances for future detections. Although the redshift distribution of FRBs is unknown, it is optimistic to gain sufficient high-redshift FRBs with the Five-hundred-meter Aperture Spherical radio Telescope (FAST) \citep{2018ApJ...867L..21Z} and future Square Kilometer Array (SKA)  \citep{2016JCAP...05..004F}. If high-redshift FRBs are observed with accurate redshift information from optical/infrared band, we can extend the $DM_{\rm IGM}-z$ relation to the epoch of reionization (EoR). The relation can constrain the parameters in reionization model and the ionized fraction precisely.

\begin{figure}[htb!]
			    \centering
	\includegraphics[width=0.5\textwidth]{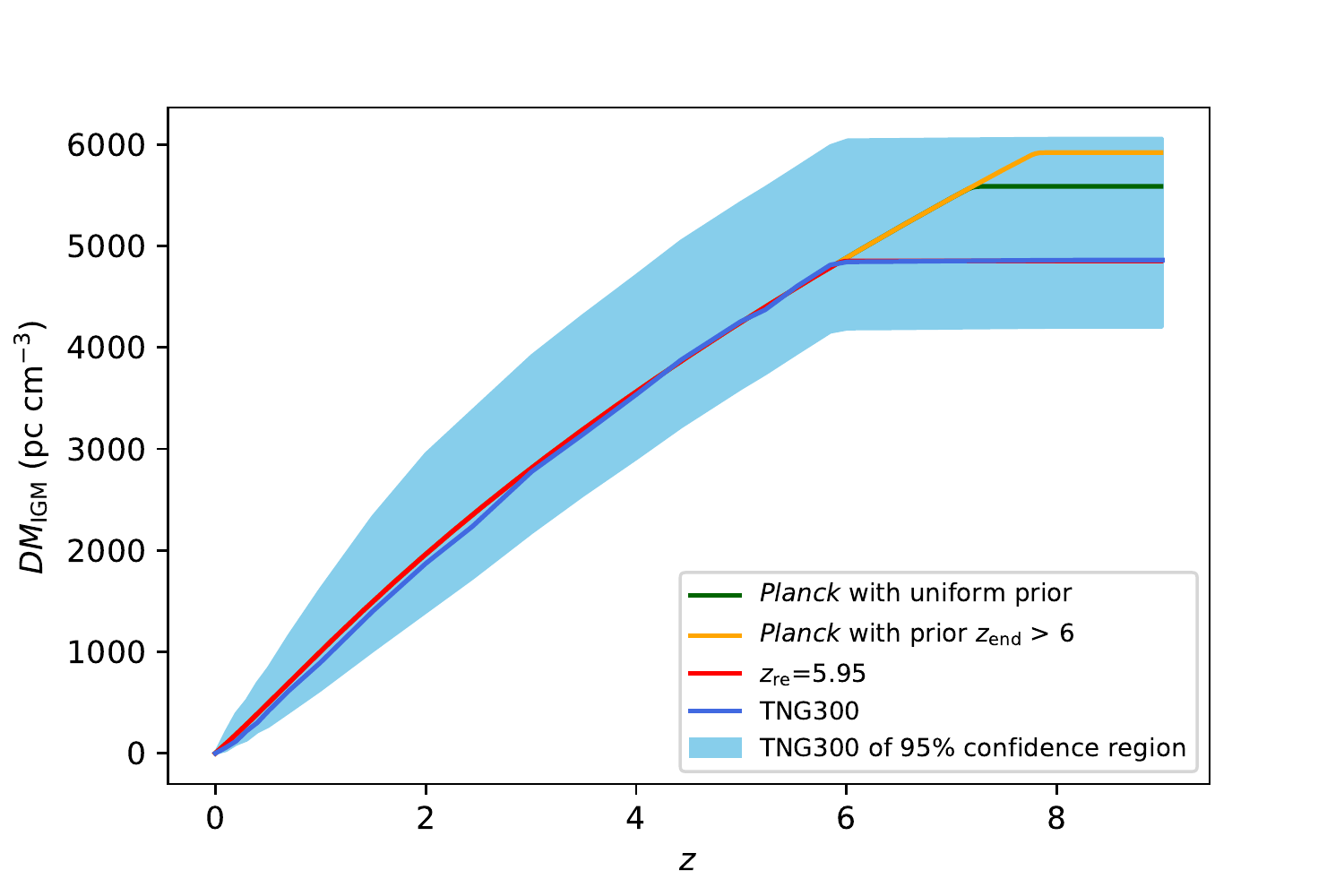}
	\caption{The blue line is our result and the red line is the best fit with the $tanh$ model. For comparison, the value derived from \textit{Planck} data is also shown \citep{2016A&A...596A.108P}. The green line takes $z_{\rm re} = 7.2$, which is the result from a uniform prior on the redshift at which the reionization ends. The orange line is the result from the prior $z_{\rm end}\ \textgreater \ 6$ and it gives $z_{\rm re} = 7.8$.}
	\label{dm-z}
\end{figure}

The wide redshift range of IllustrisTNG gives us a chance to constrain the cosmic reionzation history. We use the \textit{tanh} model given by \citet{2008PhRvD..78b3002L} (also applied by \citet{2016A&A...596A.108P}):
\begin{equation}\label{mod}
x_{\rm e}(z)=\frac{f}{2}[1+\tanh(\frac {y_{\rm re}-y}{\Delta y})],
\end{equation}
where $x_{\rm e}$ is ionized fraction. \textit{f} is a normalized parameter for considering both hydrogen and helium and expressed as $f=1+f_{\rm He}=1+n_{\rm He}/n_{\rm H}$. Typically $f \sim 1.08$, $y=(1+z)^{3/2}$, $y_{\rm re}=(1+z_{\rm re})^{3/2}$ and $z_{\rm re}$ is defined as the redshift at which $x_{\rm e}=f/2$.  $\Delta y=1.5\sqrt{1+z_{\rm re}}\Delta  z$ and $\Delta z$ reflects the duration of reionization. In the model of \citet{2008PhRvD..78b3002L}, $\Delta z$ is a fixed value. It was estimated with an upper limit of 1.3 at 95\% confidence level in \citet{2016A&A...596A.108P} redshift-symmetric case. Then we calculate $n_{\rm e}(z)$ from
\begin{equation}
 n_{\rm e}(z)=\overline n_{\rm b}(z) X_{\rm H} x_{\rm e}(z)
\end{equation}
and
\begin{equation}
\overline n_{\rm b}(z) = \frac {\Omega_{\rm b} \rho_{\rm cr,0}(1+z)^3}{m_{\rm p}},
\end{equation}
where $\rho_{\rm cr,0}$ is the critical density of universe. After taking $n_{\rm e}(z)$ into equation (\ref{dm_final}), the theoretical value of $DM_{\rm IGM}$ can be obtained in this cosmic reionization model. Figure \ref{dm-z} gives the value of $DM_{\rm IGM}$ as a function of redshift up to $z\sim9$. The blue line shows the derived $DM_{\rm IGM}$ from TNG300 simulation with 95\% confidence region (blue region). Cosmic reionization affects both CMB power spectrum and kinematic Sunyaev-Zeldovich (kSZ) effect. We display the results from \textit{Planck} data combined with the kSZ effect \citep{2016A&A...596A.108P} with the \textit{tanh} model (see equation (\ref{mod})) for comparison as well. The green line takes $z_{\rm re} = 7.2$, which is the result from a uniform prior on the redshift at which the reionization ends ($z_{\rm end}$). The orange line is the result from the prior $z_{\rm end}\ \textgreater \ 6$ and it gives $z_{\rm re} = 7.8$. The least-square-method fit (red line) indicates a very fast reionization process ($\Delta z = 0.05$ ) at $z_{\rm re}=5.95$ with the \textit{tanh} model, which is also compatible with the model used by IllustrisTNG \citep{2009ApJ...703.1416F}\footnote{Dec 2011 version, \url{https://galaxies.northwestern.edu/uvb-fg09/}}. Therefore, high-redshift FRBs are promising probes of the cosmic reionization history.

\begin{figure}[htb!]
	\centering
	\includegraphics[width=0.5\textwidth]{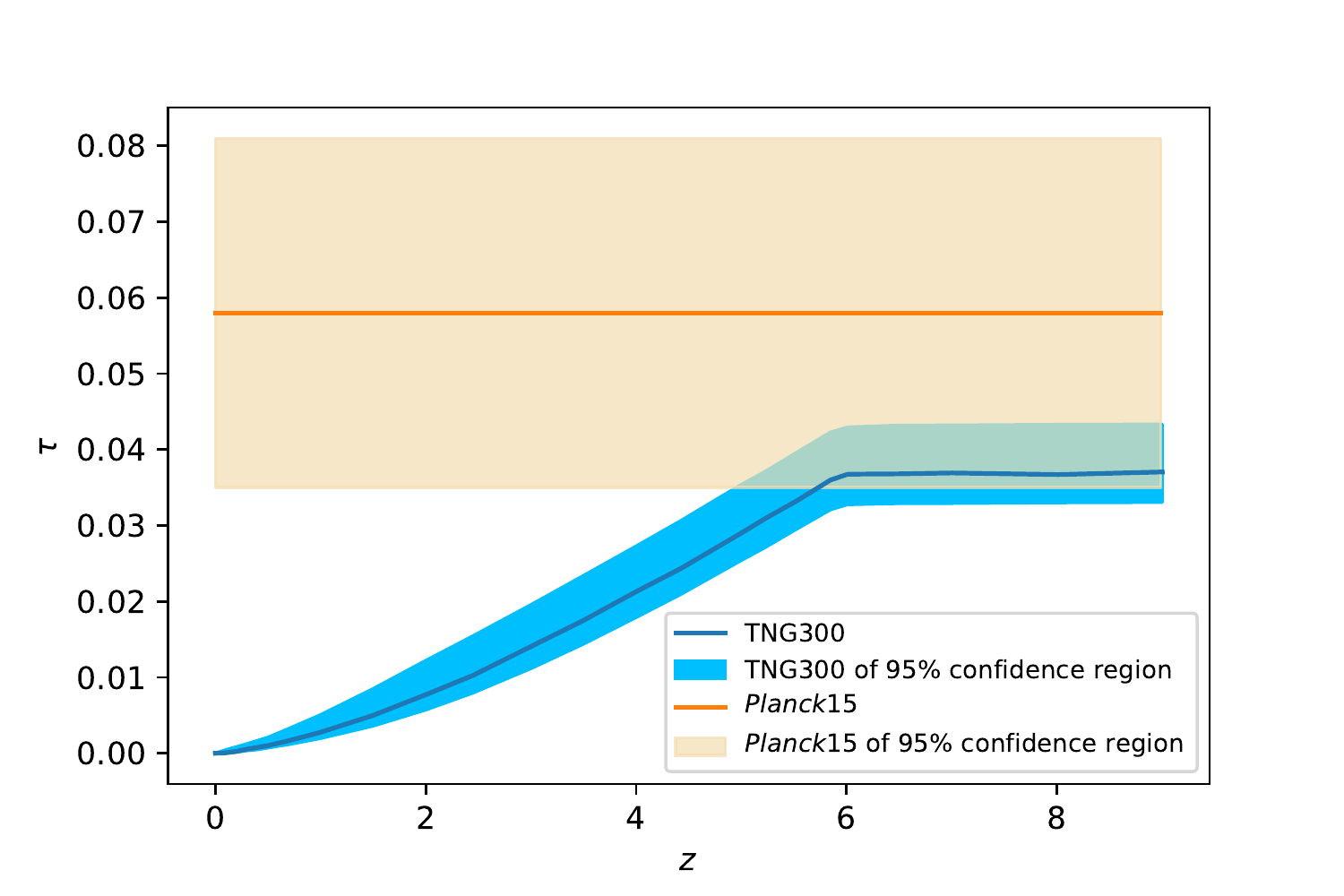}
	\caption{The blue line and shade region is our result with 95\% confidence region. The orange line and region is $Planck$ result with 95\% confidence region \citep{2016A&A...596A.108P}, which gives $\tau=0.058 \ \pm \ 0.023$.}
	\label{tau}
\end{figure}

Meanwhile, we calculate the optical depth $\tau(z)$ of CMB contributed by IGM from
\begin{equation}
\tau(z)=\int_{0}^{z} \sigma_{\rm T} n_{\rm e}(z^\prime) {{\rm d}l_{\rm prop}},
\end{equation}
where $\sigma_{\rm T}=6.25\times 10^{-25}$ cm$^2$ is the Thompson scattering cross section. The measurement of $\tau$ is changing for different instruments and we are curious whether FRBs may help. Using the electron number density derived from TNG300 simulation, we find that the total optical depth $\tau(z>6)=0.037^{+0.006}_{-0.004}$ with 95\% confidence level. Figure \ref{tau} shows the value of $\tau(z)$ as a function of redshfit from TNG300 simulation (blue line). The ionized electron fraction drops to zero before reionization so the optical depth saturates at high redshifts \citep{2016JCAP...05..004F}. The saturation value can be detected by high redshift FRBs which have no connection with CMB. The optical depth $\tau=0.058\ \pm \ 0.023$ \ \ (95\% confidence level) from Planck result is shown as the orange region \citep{2016A&A...596A.108P}. Considering the uncertainties, we find these two results are consistent with each other. Therefore, the DM of high-redshift FRBs provides an independent way to measure the optical depth of CMB, which can tightly constrain the cosmic reionization history. However, we must state that the resolution of \textit{Planck} is 12 degrees which is much larger than that of FRBs. Comparing the two results with different resolution directly may be kind of inappropriate, however there is no need to decrease the resolution of FRBs.

\section{Estimating the redshifts of non-localized FRBs} \label{sec:est}
As mentioned before, there are only ten localized FRBs \citep{2017Natur.541...58C, 2019Sci...365..565B, 2019Natur.572..352R, 2019Sci...366..231P, 2020Natur.577..190M, 2020Natur.581..391M}, which means the redshifts for most FRBs are unknown. Assuming that the accurate $DM_{\rm IGM}$ is known, we check whether the redshift of FRB can be precisely derived from the $z-DM_{\rm IGM}$ relation.

\begin{figure}[htb!]
        \centering
	\includegraphics[width=0.5\textwidth]{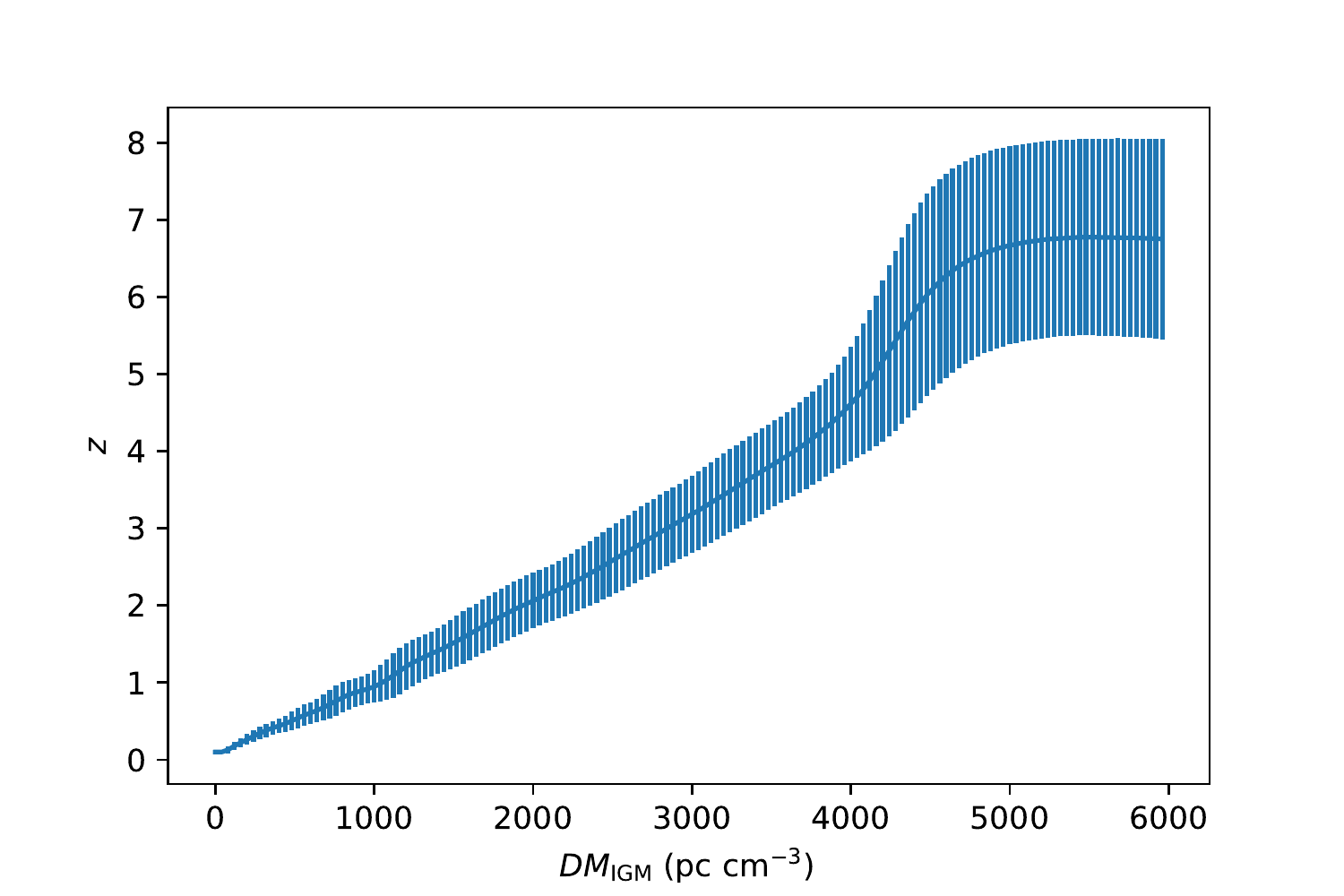}
	\caption{The $z-DM_{\rm IGM}$ relation. The short blue lines show one standard deviation of each bin.}
	\label{eor}
\end{figure}

As mentioned in Section \ref{sec:method} and Section \ref{sec:result}, we get $23 \times 10,000,000$ $DM_{\rm IGM}-z$ data points (FRBs) in total (24 snapshots and 10,000,000 combinations). We divide the data points whose $DM_{\rm IGM}$ is between 0 and 6000 pc cm$^{-3}$ into 150 bins (40 pc cm$^{-3}$ in each bin) and calculate the mean redshift of these FRBs in each bin. The $z-DM_{\rm IGM}$ relation is shown in Figure \ref{eor}. The standard deviation of derived redshift is 0.74 at the pseudo-redshfit 4.61 ($DM_{\rm IGM}=4000$ pc cm$^{-3}$), which means the relative error is 16.1\%. The small redshift standard deviations of FRBs whose $DM_{\rm IGM}$ are less than 4000 pc cm$^{-3}$ show a good prospect for calculating pseudo-redshfits of non-localized FRBs.

We only take full snapshots at low redshifts, which means the redshifts of our simulated FRBs are restricted to some fixed redshifts such as 0.1, 0.2, 0.3, 0.4, 0.5, 0.7, 1 and 1.5. However, the standard deviation should not be larger when including mini snapshots because the standard deviation does not change for different sampling methods. What's more, the 40 pc cm$^{-3}$ bin width can be taken as the systematic error during the estimation of $DM_{\rm halo}$, $DM_{\rm host}$ and $DM_{\rm source}$, which means it is even optimistic to derive the pseudo-redshift with a `not very accurate' $DM_{\rm IGM}$.

\section{Conclusions}\label{sec:con}
In this work, we derive $DM_{\rm IGM}$ of FRBs at the redshift range of $0\ \textless \ z\ \textless \ 9$ from IllustrisTNG simulation. We obtain $DM_{\rm IGM} = 892^{+721}_{-270}$ pc cm$^{-3}$ at $z=1$. The $DM_{\rm IGM}$ value of localized FRBs are consistent with the derived $DM_{\rm IGM}-z$ relation. At high redshifts $z>5$, we show the scenario of probing the cosmic reionization history with FRBs. The $tanh$ reionization model is used to fit the derived $DM_{\rm IGM}-z$ relation at high redshifts. We find the reionization of IllustrisTNG universe occurs quickly at $z = 5.95$. This reionization model is compatible with the theoretical model used by the IllustrisTNG simulation \citep{2009ApJ...703.1416F}. The optical depth of CMB is also derived from the IllustrisTNG simulation, which is consistent with that from \citet{2016A&A...594A..13P}. We try to limit the redshifts of non-localized FRBs with their $DM_{\rm IGM}$. The standard deviation of pseudo-redshifts for non-localized FRBs is 16.1\% for $DM_{\rm IGM}=4000$ pc cm$^{-3}$.

The highest DM of observed FRB is $2596.1 \pm 0.3$ pc cm$^{-3}$ and its nominal redshift is 2.1 \citep{2018MNRAS.475.1427B,2018MNRAS.478.2046C}. It is predictable that there will be two orders of magnitude more FRBs in the next few years \citep{2018NatAs...2..865K}. According to \citet{2016JCAP...05..004F} and \citet{2018ApJ...867L..21Z}, FAST and SKA will have enough capability to detect FRBs out to $z = 14\sim15$. Therefore, FRBs will be a powerful and independent probe of the universe during epoch of reionization besides hydrogen 21-cm line. The future large sample of FRBs can be used to test our result about estimating redshifts for non-localized FRBs. What's more, most of previous works only considered the mean value of $DM_{\rm IGM}$ (equation \ref{DM_{igm}}) for cosmological constraints, while the scatter of it (Figure \ref{dis}) which can degrade the cosmological constraints was not handled properly. Since IllustrisTNG shows a certain shape of the scatter, more authentic distributions of $DM_{\rm IGM}$ and $DM_{\rm host}$ can be used to test the FRB's capability of constraining cosmological parameters, such as $\Omega_{\rm m}$ and $\Omega_{\rm \Lambda}$.

\acknowledgements
We thank the anonymous referee for valuable comments. We thank Zhimei Tan, Lingrui Lin and Yichen Sun for helpful discussions and thank Dylan Nelson for his theoretical and technical help.
This work is supported by the National Natural Science Foundation of China (grant U1831207).


\end{document}